\documentclass[osajnl,twocolumn,showpacs,superscriptaddress,10pt]{revtex4-1}

\usepackage{amsmath,amssymb,graphicx}

\begin{document}

\title{Quartz-enhanced photoacoustic spectroscopy as a platform for non-invasive trace gas analyser targeting breath analysis}

\author{Jan C. Petersen}
\affiliation{Danish Fundamental Metrology, Matematiktorvet 307, DK-2800 Kgs. Lyngby, Denmark}

\author{Laurent Lamard}
\affiliation{Laserspec BVBA, 15 rue Trieux Scieurs, B-5020 Malonne, Belgium}

\author{Yuyang Feng}
\affiliation{COPAC ApS, Diplomvej 381, Scion science park, DK-2800, Kgs. Lyngby, Denmark}

\author{Jeff-F. Focant}
\affiliation{University of Liege, Allee du 6 août, B6, 4000 Liege, Belgium}

\author{Andre peremans}
\affiliation{Laserspec BVBA, 15 rue Trieux Scieurs, B-5020 Malonne, Belgium}

\author{Mikael Lassen}\email{Corresponding author: ml@dfm.dk}
\affiliation{Danish Fundamental Metrology, Matematiktorvet 307, DK-2800 Kgs. Lyngby, Denmark}

\begin{abstract}
An innovative and novel quartz-enhanced photoacoustic spectroscopy (QEPAS) sensor for highly sensitive and selective breath gas analysis is introduced. The QEPAS sensor consists of two acoustically coupled micro-resonators (mR) with an off-axis 20 kHz quartz tuning fork (QTF). The complete acoustically coupled mR system is optimized based on finite element simulations and experimentally verified.  Due to the very low fabrication costs the QEPAS sensor presents a clear breakthrough in the field of photoacoustic spectroscopy by introducing novel disposable gas chambers in order to avoid cleaning after each test. The QEPAS sensor is pumped resonantly by a nanosecond pulsed single-mode mid-infrared optical parametric oscillator (MIR OPO). Spectroscopic measurements of methane and methanol in the 3.1 $\mu$m to 3.7 $\mu$m wavelength region is conducted. Demonstrating a resolution bandwidth of 1 cm$^{-1}$. An Allan deviation analysis shows that the detection limit at optimum integration time for the QEPAS sensor is 32 ppbv@190s for methane and that the background noise is solely due to the thermal noise of the QTF.  Spectra of both individual molecules as well as mixtures of molecules were measured and analyzed.  The molecules are representative of exhaled breath gasses that are bio-markers for medical diagnostics.
\end{abstract}

\maketitle

\section{INTRODUCTION}
\label{sec:intro}  % \label{} allows reference to this section

   \begin{figure} [ht]
   \begin{center}
   \begin{tabular}{c} %% tabular useful for creating an array of images
   \includegraphics[width=\linewidth]{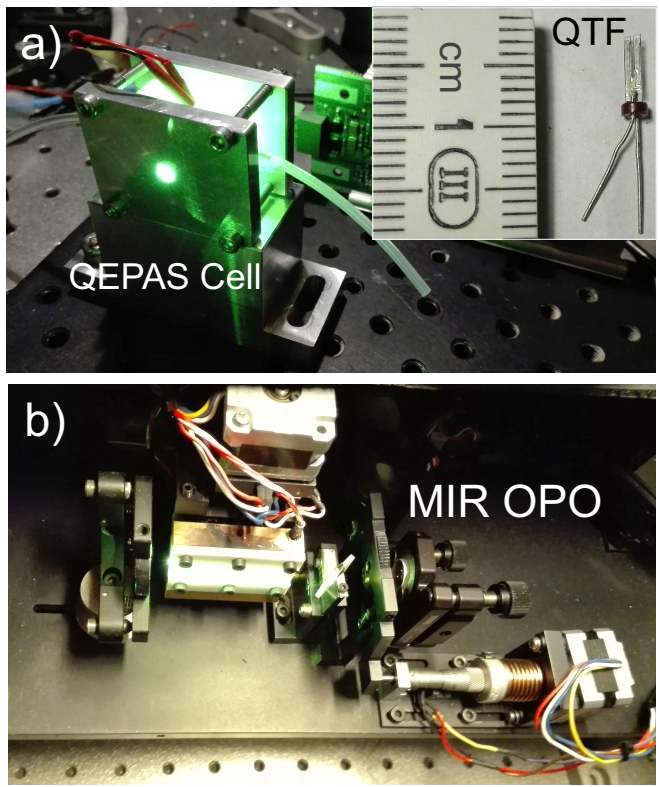}
   \end{tabular}
   \end{center}
   \caption[example]
%>>>> use \label inside caption to get Fig. number with \ref{}
   { \label{fig1} The two main components of the novel non-invasive trace gas analyser. a)  The QEPAS sensor. Inlet: The size of the quartz tuning fork (QTF). Due to the very low fabrication costs the QEPAS sensor presents a clear breakthrough in the field of photoacoustic spectroscopy by introducing novel disposable gas chambers in order to avoid cleaning after each test.  b) The diode pumped nano second MIR OPO.  OPO systems are desirable due to their reliability, tunability and ease of operation. The maximum wavelength ranges for the signal and idler are 1.4 to 1.7 $\mu$m and 2.8 to 4.8 $\mu$m, respectively. }
   \end{figure}

Breath analysis enables the diagnosis of a specific disease by analysing the changes in breath gas composition. Tests revealing inflammatory processes in e.g. the lungs, neonatal jaundice and allergies based on inorganic gases concentrations is currently marketed. Currently there is also interest in VOC measurement, as they can be used to diagnose multiple metabolic disorders. Today breath gas analysis is typically carried out using gas chromatography and mass spectrometers. Recent research demonstrated such an approach can help differentiate between clinical patients and controls with various stages of cancer \cite{Buszewski2012,Lourenco2014,Lindinger1998}. Unfortunately cost and speed has limited its use mainly to research. The development of a cost effective and portable platform opens the possibility of promoting breath gas analysis as a very promising in-situ screening method. Trace gas and breath gas analysis measurements in the MIR and NIR wavelength regions are particular important due to the presence of strong ro-vibrational bands of most molecules resulting in high sensitivity \cite{Hodgkinson2013,Sigrist2003,Sigrist2008}. Photoacoustic spectroscopy (PAS) is a very promising method due to its ease of use, relatively low cost and the capability of allowing trace gas measurements at the sub-ppb level \cite{Spagnolo2012,Peltola2015,Szabo2013,Lassen2014}. The photoacoustic (PA) technique is based on the detection of sound waves that are generated due to absorption of modulated optical radiation. Quartz tuning forks (QTFs) have shown great potential as sound transducers for PAS and have been increasingly applied to selective and sensitive detection of trace gases since its introduction in 2002 \cite{Kosterev2002}. Standard low cost QTFs with resonance frequencies at 32.7 kHz are typically used as sensors, however, also custom made QTFs have been reported \cite{Patimisco2014,Spagnolo2015,Zheng2016,Lassen2016qepas,Kohring2015,Yi2014,Dong2014}. The PA signal is proportional to the Q-factor: $S \propto Q P \alpha /f_0$, where $P$ is the optical power, $\alpha$ is the molecular absorption coefficient and $f_0$ is the resonant frequency of the QTF.  The stiffness of quartz provides an efficient means of confining the acoustic energy in the prongs of the QTF, resulting in large quality factors (Q-factors) of the order 100000 in vacuum and 3-8000 at atmospherical pressures \cite{Dong2010}. This enables therefore  detection of very weak PA excitation with very small gas volumes.

The main objective of this research is to develop, test and demonstrate a novel non-invasive trace gas analyser platform targeting the major market opportunity of medical diagnostics based on breath analysis. The device will be capable of detecting early stages of diseases almost instantaneously providing data for potential diagnostics. The  sensor will be non-invasive and have fast screening capability thus significantly improve early diseases diagnosis for the benefit of patients. We report on a novel QEPAS sensor design with two mR tubes, an off-axix acoustically coupled mR tube, where a 20 kHz QTF is placed \cite{Lassen2016qepas}. The off-axis mR system has several advantages compared with traditional on-axis QEPAS systems. It makes optical alignment easy and thus background free measurements are easier to achieve. We find that the resulting background noise signal is solely due to the QTF thermal noise. Another advantage is that it allows mechanical protection of the QTF, since only the top of the QTF casing needs to be opened. The QEPAS measurements are conducted with a nanosecond pulsed single-mode MIR OPO. The resonance of the QTF is excited resonantly by setting the repetition rate of the OPO to 20 kHz. Figure~\ref{fig1}a) shows a picture of the QEPAS sensor.The insert in Fig~\ref{fig1}a) shows the size of the quartz tuning fork (QTF) used.  The QTF is 5 mm long and the disctance between prongs are 0.4 mm Figure~\ref{fig1}b) shows a picture of the nanosecond pulsed single-mode MIR OPO developed by laserspec.

\section{Experimental Setup}

For highly sensitive and selective trace-gas sensing it is desirable to have high power sources with large wavelength tunability in the mid-infrared (MIR) region, where most molecules have strong vibrational transitions (fingerprint region) \cite{Sigrist2008}. A number of different light sources (QCLs, LEDs, DFBs, OPOs and more) have been reported used in QEPAS experiments \cite{Kohring2015,Spagnolo2015,Ma2013,Yi2014,Patimisco2014,Dong2010}. OPOs seem to be the optimal choice for providing large wavelength tunability, high energy, molecular selectivity and cost-effective device for the generation of infrared light in the 1.5 to 5 $\mu$m spectral range \cite{Petrov2015}. Therefore, a pulsed single mode mid-infrared (MIR) OPO has been developed. The pulse repetition rate is matched to the resonance frequency of the acoustic resonance of the QTF and the QEPAS sensor. The OPO is based on a 50 mm long PPLN nonlinear crystal with a fanned-out structure. The PPLN is placed inside a single-resonant cavity. The OPO is pumped at 1064 nm with diode-pumped nanosecond laser with an average output power reaching up to 27W. The Q-switch repetition rate can be changed continuously from 10 kHz to 80 kHz and the pulse duration from 7 to 50 ns. The maximum wavelength ranges for the signal and idler are 1.4 to 1.7 $\mu$m and 2.8 to 4.8 $\mu$m, respectively, and with average output power ranging up to 4 W. The tunability of the OPO is achieved by the vertical translation of the fanned-out PPLN crystal, while the temperature of the crystal is kept at 30$^\circ$C. The system is kept single-mode using a 50$\mu$m thick etalon plate. This provides a bandwidth of around 1 cm$^{-1}$. A spectrometer is integrated in the system and the position of the stepping motors of the crystal mount, the etalon plate and the grating are controlled by a computer. In the present work the OPO is optimized for operation in the spectral region between 3.1 $\mu$m and 3.8$\mu$m and with an average output power of approximately 400 mW and pulse durations of $\sim$ 18 nanoseconds at a repetition rate of 19.99 kHz, thus matching the QTF and mR tubes.

   \begin{figure} [ht]
   \begin{center}
   \begin{tabular}{c} %% tabular useful for creating an array of images
   \includegraphics[width=\linewidth]{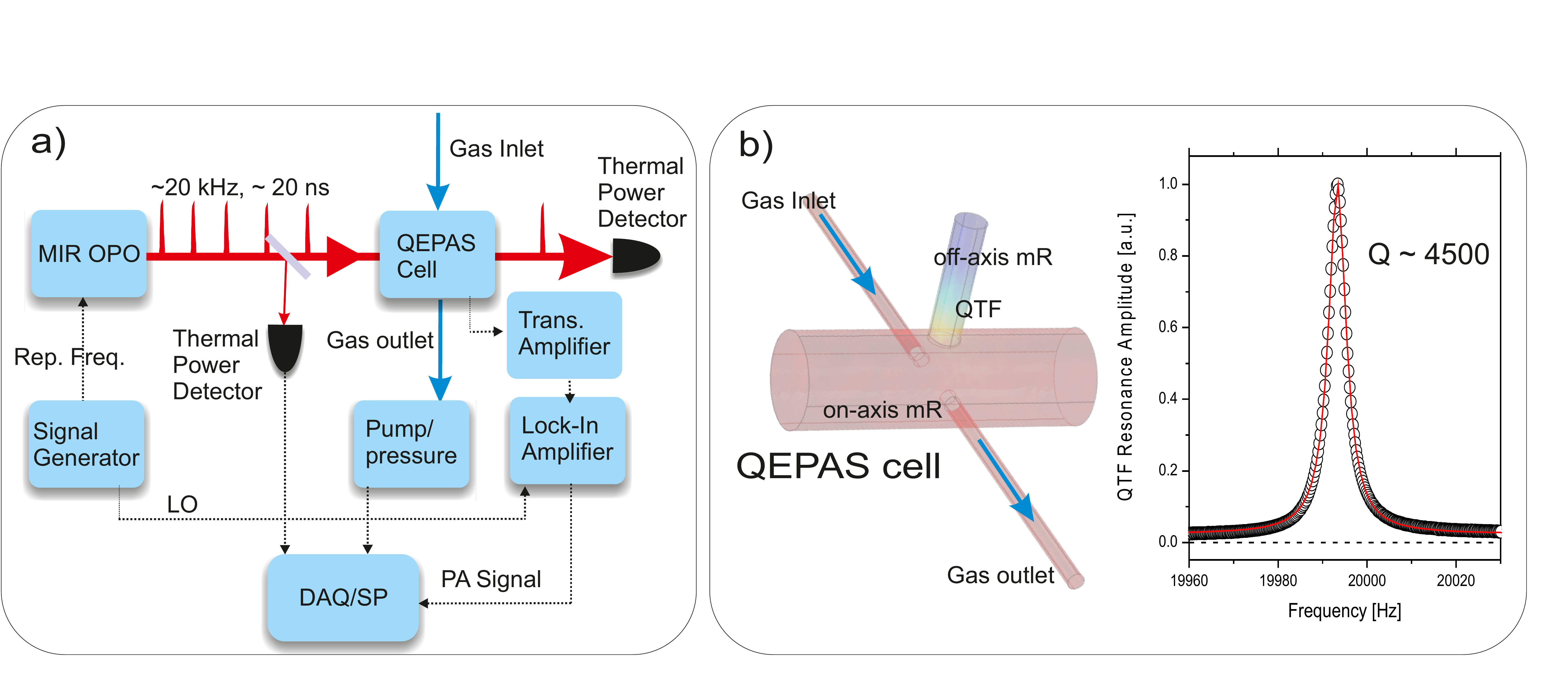}
   \end{tabular}
   \end{center}
   \caption[example]
%>>>> use \label inside caption to get Fig. number with \ref{}
   { \label{fig2}
a) Block diagram of the important parts of the experimental setup. Lenses and mirrors are not shown. The experimental setup includes a nanosecond MIR OPO tunable between 3.1 and 3.8 $\mu$m, a novel mR system, a 20 kHz QTF, thermal detectors for monitoring the optical power, a signal generator for controlling the OPO repetition rate, pressure and temperature controllers for the QEPAS sensor and a lock-in amplifier connected to a 12 bit DAQ card. LO: Local oscillator, SP: Signal Processing. b) Layout of the QEPAS cell with the two acoustic coupled mircro resonantors (mRs). Normalized frequency response of the QTF. Measured Q-factor $\sim$ 4500. }
   \end{figure}

The mR cell was constructed based on finite element simulations in high-density PTFE as depicted in Figure \ref{fig2}b) \cite{Baumann2007,Lassen2015}. The dimensions and design of the optimal design is shown in figure \ref{fig2}b). The on-axis mR (absorption tube) is 9.5 mm-long with 1.5 mm inner diameter coupled to the off-axis mR with a 4.1 mm long tube with 1 mm inner diameter, where the QTF is placed. The gas inlet/outlet tubes have a 0.5 mm inner diameter. This design seems to yield the best acoustic coupling to the QTF at 20 kHz. The PTFE material is used for several reasons. The PTFE walls decouples the in-phase background absorption signal from the PA gas signal due to thermal diffusion effects \cite{Lassen2014}. The MIR light beam enters and exits the on-axis mR via uncoated 3 mm thick calcium fluoride windows. For optimal transmission of the MIR light beam several lenses and mirrors are used (not shown in Figure \ref{fig1}a)). The measured optical transmission through the cell is 90$\%$. The QTF is placed inside the off-axis mR tube. The pressure is kept at atmospheric pressure at all times and the temperature of the cell is kept at 25$^\circ$C. This temperature has been chosen to optimize the acoustic coupling of PA signal generated in the on-axis tube to the off-axis QTF. The current signal from the QTF is first amplified by a transimpedance amplifier and then amplified by a pre-amplifier with a 1 kHz bandpass filter at 20 kHz before being processed with a lock-in amplifier and finally digitized with a 12 bit DAQ card.

\section{Measurements}

   \begin{figure} [ht]
   \begin{center}
   \begin{tabular}{c} %% tabular useful for creating an array of images
   \includegraphics[width=\linewidth]{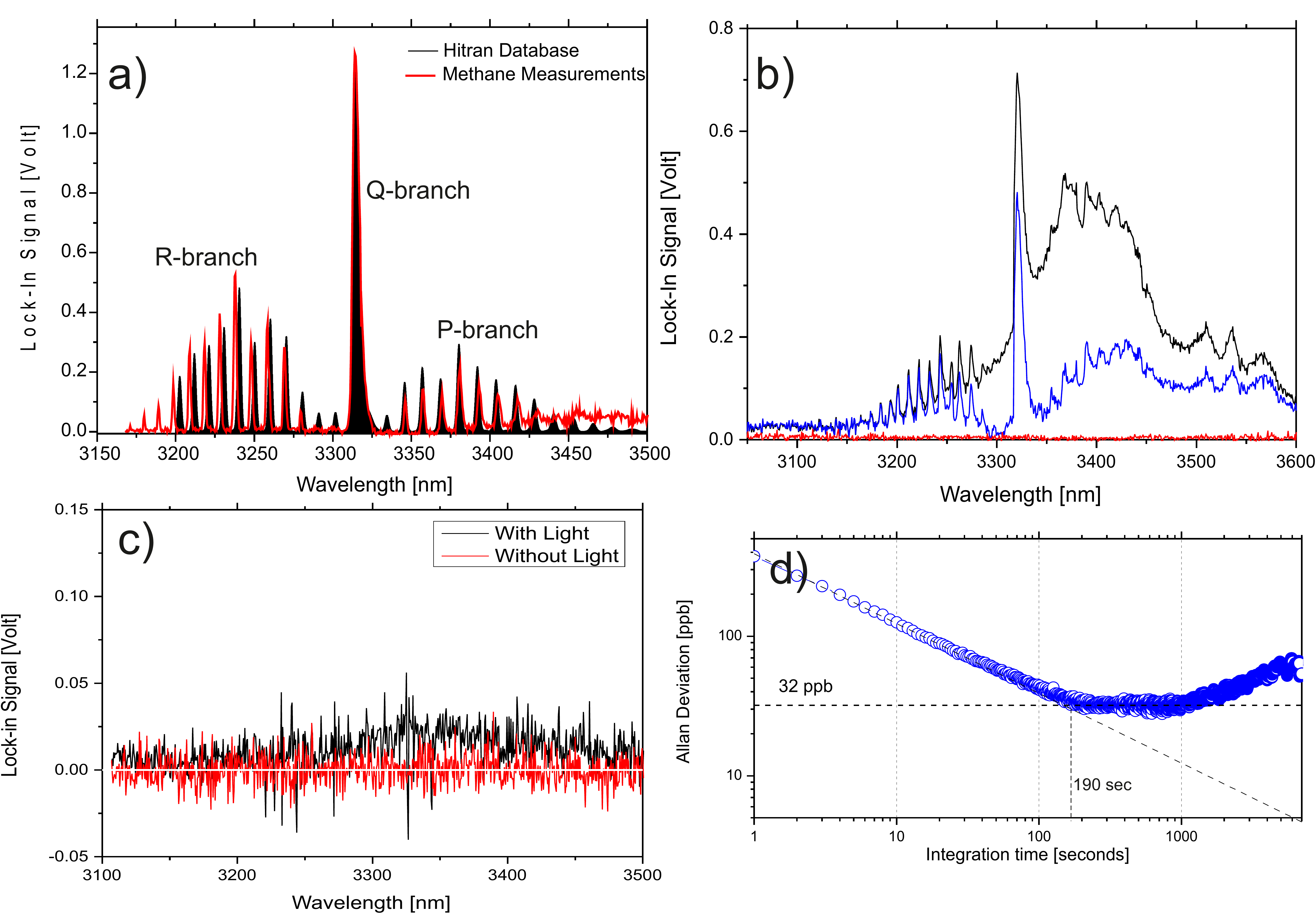}
   \end{tabular}
   \end{center}
   \caption[example]
%>>>> use \label inside caption to get Fig. number with \ref{}
   { \label{fig3}
a) The red curve shows the measured methane spectrum with 100 ppm concentration and at 1 atm pressure. The black curve shows data from the Hitran database. b) Gas mixture between methane and methanol.  c) The background noise with and without light.  d) Allan deviation analysis of the lock-in signal with 10 seconds integration and measured at 3.314 $\mu$m. . Detection sensitivity at optimum integration time is 32 ppbv@190s.}
   \end{figure}

The experiments were performed by excitation of molecular ro-vibrational transitions of methane and methanol in synthetic dry air in the 3.1-3.7 $\mu$m wavelength region. Both methane and methanol detection is of considerable interest since it is considered as a potential biomarker for different gut and stomach inflammatory diseases and colorectal and lung cancer \cite{Lourenco2014}. Figure \ref{fig3}a) shows the measured spectrum of methane (CH$_4$) with a 100 ppm concentration in synthetic air. We clearly see the R-, Q- and P-branch of methane. The Q-branch has a peak value of 1.28V. The data shown in Figure \ref{fig3}a) and b) were processed with a lock-in amplifier with a time constant of 300 ms, this corresponds to the energy accumulation time of the QTF at atmospheric pressure. The wavelength of the OPO was changed in steps of 0.5 nm. Figure \ref{fig3}b) shows the measured spectrum of a mixture between methane (CH$_4$) and methanol (CH$_3$OH). We clearly see that our QEPAS sensor can differentiate methane from methanol. However since the PAS technique is not an absolute technique calibration is required against a known sample concentration \cite{Lassen2017}. We can therefore use the measured methane signal for the calibration of the QEPAS sensor and in this case to estimate the concentration of unknown amount of methanol. This gives us an estimated concentrations of approximately 35$\pm$5 ppm and 15$\pm$5 ppm of methanol for the two different curves shown in Figure \ref{fig3}b), respectively. Figure \ref{fig3}c) shows the measured background signal with and without light. The background noise was measured by flushing the cell with pure air followed by a wavelength scan. We find that the background noise, with and without light, has more or less the same magnitude and standard deviation. However due to non proper cleaning of the cell we find a small difference between the background signals, which we believe is because of a small methane concentration of approximately 1-2 ppm, since the peak signals fits very well with the Q-branch of methane. We conclude that the background noise is not affected by stray light or cell absorption noise.

In order for the non-invasive trace gas analyser platform to be practical having a small resolution bandwidth and high tunability is not sufficient the sensor also needs high sensitivity. Since the monitoring of a disease requires a certain change in the amount of the exhaled gas associated with the disease. If the biomarker is not present in the natural exhaled breath from a person the required minimum amount detectable for a diagnosis has to be estimated. If a biomarker is naturally present the minimum re-quired change for making a diagnosis has to be estimated. This knowledge will provide information about the required sensitivity and resolution of the breath analyser. We have already seen from the background signal measurements that the system, with only 300 ms of integration time, can measure methane concentrations down to 1 ppm. However the optimum integration time and detection sensitivity are therefore determined using an Allan deviation analysis. The OPO was locked to the Q-branch absorption line peak while the methane concentration is maintained constant. The Allan deviation is based on data consisting of 100k points and was recorded over 12 minutes. Figure \ref{fig3}d) shows the Allan deviation analysis for the QEPAS data processed with a lock-in amplifier with a integration time constant of 10 seconds. The analysis shows that the detection sensitivity at optimum integration time is 32 ppbv@190s for methane measured at 3.314 $\mu$m. This means that white noise remains the dominant noise source for 190 s. This is where the ultimate detection limit is reached and after 1000 s the instrumental drift starts dominating.

\section{Conclusion}
In conclusion, we have demonstrated a novel QEPAS configuration using acoustic coupled micro-resonators with an off-axis 20 kHz quartz tuning fork (QTF). The QEPAS system is pumped resonantly by a nanosecond pulsed single-mode MIR OPO. To our knowledge these spectroscopic measurements are the first to combine QEPAS with a high power widely tunable nanosecond pulsed MIR OPO. Different spectral features of methane is resolved and the R-, Q- and P-branchs are clearly identified. From the comparison of the methane spectra with the Hitran spectra we conclude that the QEPAS instrument resolution bandwidth is approximately 1 cm$^{-1}$. By applying optimum integration time the sensitivity can be increased to 32 ppbv@190s. We believe that the tunability and sensitivity demonstrated here is sufficient to make the QEPAS sensor system very useful for environmental, industrial, and biological monitoring, where multiple gaseous pollutants and aerosols need to be monitored simultaneously. Further improvement of the QEPAS sensor will be made to make the complete system more compact and robust in order perform in situ monitoring outside the safe environment of the laboratory. The cost effectiveness and portability of the QEPAS platform opens the possibility of promoting breath analysis as a fast and cost effective screening method. This makes the QEPAS sensor a very competitive alternative to gas chromatography and mass spectrometry and likely to be one of the devices of choice in future medical diagnostics.

We acknowledge the financial support from EUREKA (Eurostars program: E9117NxPAS) and the Danish Agency for Science Technology and Innovation.


\begin{thebibliography}{99}

\bibitem{Buszewski2012} Buszewsk, B.i, Ligor, T., Jezierski, T., Wenda-Piesik, A., Walczak, M., and  Rudnicka, J., "Identification of volatile lung cancer markers by gas chromatography-mass spectrometry: Comparison with discrimination by canines", Anal. Bioanal. Chem., \textbf{404}, 141--146 (2012)

\bibitem{Lindinger1998} Lindinger,  W.,  Hansel, A., and  Jordan, A., "On-line monitoring of volatile organic compounds at pptv levels by means of proton-transfer-reaction mass spectrometry (PTR-MS)—Medical applications, food control and environmental research",  Int J Mass Spectrom \textbf{173} 191--241 (1998).

\bibitem{Lourenco2014}  Lourenço, C.,  and  Turner, C., "Breath analysis in disease diagnosis: methodological considerations and applications", Metabolites \textbf{4}, 465--498 (2014).

\bibitem{Hodgkinson2013} Hodgkinson,  J. and Tatam,  R. P., "Optical gas sensing: a review", Meas. Sci. Technol., \textbf{24}, 012004 (2013).

\bibitem{Sigrist2003}  Sigrist, M. W., "Trace gas monitoring by laser photoacoustic spectroscopy and related techniques ", Rev. Sci. Instrum. \textbf{74(1)}, 486--490 (2003).

\bibitem{Sigrist2008}  Sigrist, M. W., Bartlome,  R., Marinov,  D.,  Rey J. M., Vogle,r  D. E., and Wächter, H., "Trace gas monitoring with infrared laser-based detection schemes", Appl. Phys. B \textbf{90}, 289--300 (2008).

\bibitem{Spagnolo2012}  Spagnol,o V., Patimisco,  P., Borri,  S.,  Scamarcio, G., Bernacki,  B., and Kriesel,  J., "Part-per-trillion level SF6 detection using a quartz enhanced photoacoustic spectroscopy-based sensor with single-mode fiber-coupled quantum cascade laser excitation", Opt. Lett. \textbf{37}, 4461--4463 (2012).

\bibitem{Peltola2015}   Peltola,  J., Hieta,  T., and  Vainio, M., "Parts-per-trillion-level detection of nitrogen dioxide by cantilever-enhanced photo-acoustic spectroscopy", Opt. Lett. \textbf{40}, 2933--2936 (2015).

\bibitem{Szabo2013}   Szabó,  A. , Mohacsi,  A., Gulyas,  G.,  Bozoki, Z., and Szabo,  G.,  "In situ and wide range quantification of hydrogen sulfide in industrial gases by means of photoacoustic spectroscopy", Meas. Sci. Technol. \textbf{24}(6), 065501 (2013).

\bibitem{Lassen2014} Lassen, M., Balslev-Clausen,  D., Brusch,  A., and Petersen,  J. C. , "A versatile integrating sphere based photoacoustic sensor for trace gas monitoring",  Opt. Express \textbf{22}, 11660--11669 (2014).

\bibitem{Kosterev2002}  Kosterev, A.A.,  Bakhirkin, Y.A.,  Curl, R.F., and  Tittel, F.K., "Quartz-enhanced photoacoustic spectroscopy", Opt. Lett. \textbf{27}, 1902--1904 (2002).

\bibitem{Zheng2016} Zheng,  H., et al., "Single-tube on-beam quartz-enhanced photoacoustic spectroscopy", Opt. Lett. \textbf{41}, 978--981 (2016).

\bibitem{Dong2014}    L. Dong, et al., "Double acoustic microresonator quartz-enhanced photoacoustic spectroscopy",  Opt. Lett. \textbf{39}, 2479--2482 (2014).

\bibitem{Lassen2016qepas} Lassen,  M., Lamard,  L.,  Feng,  Y.,  Peremans,  A.,  and Petersen, J. C. Petersen, "Off-axis quartz-enhanced photoacoustic spectroscopy using a pulsed nanosecond mid-infrared optical parametric oscillator," Opt. Lett. \textbf{41}, 4118--4121 (2016).

\bibitem{Patimisco2014} Patimisco, P.,   Scamarcio, G.,  Tittel,  F. K.,  and Spagnolo,  V., "Quartz-Enhanced Photoacoustic Spectroscopy: A Review, Sensors", Sensors \textbf{14}, 6165--6206 (2014).

\bibitem{Kohring2015} Köhring, M. ,  Böttger,  S.,  Willer,  U.,  and Schade, W. , "LED-absorption-QEPAS sensor for biogas plants", Sensors, \textbf{15}, 12092--12102 (2015).

\bibitem{Spagnolo2015} Spagnolo,  V.,  Patimisco, P., Pennetta,  R.,  Sampaolo, A., Scamarcio,  G.,  Vitiello,  M. S.,  and Tittel, F. K., "THz Quartz-enhanced photoacoustic sensor for H2S trace gas detection", Opt. Express \textbf{23(6)}, 7574--7582 (2015).

\bibitem{Yi2014} Yi,  H.,  Chen, W., Vicet,  A., Cao,  Z., Gao, X. ,  Nguyen,  T.,  Jahjah, M.,  Rouillard, Y., Nähle,  L., and  Fischer,  M., "T-shape microresonator-based quartz-enhanced photoacoustic spectroscopy for ambient methane monitoring using 3.38-$\mu$m antimonide-distributed feedback laser diode", Appl. Phys. B, \textbf{116}, 423--428 (2014).

\bibitem{Dong2010} Dong,  L.,  Kosterev, A. A.,  Thomazy,  D.,  and  Tittel, F. K., "QEPAS spectrophones: design, optimization, and performance", Appl. Phys. B 100(3), 627--635 (2010).

\bibitem{Ma2013}  Ma, Y.,  Lewicki,  R., Razeghi, M., and Tittel, F. K., "QEPAS based ppb-level detection of CO and N2O using a high power CW DFB-QCL", Opt. Express \textbf{21(1)}, 1008--1019 (2013).

\bibitem{Petrov2015} Petrov, V., "Review Frequency down-conversion of solid-state laser sources to the mid-infrared spectral range using non-oxide nonlinear crystals", Progress in Quantum Electronics \textbf{42}, 1--116 (2015).

\bibitem{Lassen2015}  Lassen, M., Brusch, A.,   Balslev-Harder, D.,  and Petersen, J. C., "Phase-sensitive noise suppression in a photoacoustic sensor based on acoustic circular membrane modes", Appl. Opt. \textbf{54}, D38--D42 (2015).

\bibitem{Baumann2007} Baumann,  B.,  Wolff,  M., Kost, B., and Groninga,  H., "Finite element calculation of photoacoustic signals", Appl. Opt. \textbf{7}, 1120--1125 (2007).

\bibitem{Lassen2017} Lassen, M.,  Baslev-Harder,  D.,  Brusch,A., Nielsen, O. S., Heikens, D.,  Persijn, S. and  Petersen, J. C., "Photo-acoustic sensor for detection of oil contamination in compressed air systems," Opt. Express 25, 1806-1814 (2017)

\end{thebibliography}
\end{document}